\documentclass[twocolumn,superscriptaddress,amsmath,amssymb,showpacs,prl]{revtex4-2}

\usepackage{graphicx}
\usepackage{color}
\usepackage{makecell}
\usepackage{footnote}
\renewcommand{\phi}{\varphi}

\begin{document}

\title{Thermoelastic properties of bridgmanite using Deep Potential Molecular Dynamics}

\author{Tianqi Wan}
	\affiliation{Department of Applied Physics and Applied Mathematics, Columbia University, New York, NY 10027, USA}
 	\affiliation{Department of Physics, Xiamen University, Xiamen 361005, China}

\author{Chenxing Luo}
	\affiliation{Department of Applied Physics and Applied Mathematics, Columbia University, New York, NY 10027, USA}

\author{Yang Sun}
	\email{yangsun@xmu.edu.cn}
 	\affiliation{Department of Physics, Xiamen University, Xiamen 361005, China}
	\affiliation{Department of Physics, Iowa State University, Ames, 50011, USA}

\author{Renata M. Wentzcovitch}
	\email{rmw2150@columbia.edu}
	\affiliation{Department of Applied Physics and Applied Mathematics, Columbia University, New York, NY 10027, USA}
	\affiliation{Department of Earth and Environmental Sciences, Columbia University,New York, NY 10027, USA}
	\affiliation{Lamont–Doherty Earth Observatory, Columbia University, Palisades, NY 10964, USA}
 	\affiliation{ Data Science Institute, Columbia University, New York, NY, 10027, USA}

\date{\today}

\begin{abstract}
MgSiO$_3$-perovskite (MgPv) plays a crucial role in the Earth's lower mantle. This study combines deep-learning potential (DP) with density functional theory (DFT) to investigate the structural and elastic properties of MgPv under lower mantle conditions. To simulate complex systems, we developed a series of potentials capable of faithfully reproducing DFT calculations using different functionals, such as LDA, PBE, PBEsol, and SCAN meta-GGA functionals. The obtained predictions exhibit remarkable reliability and consistency, closely resembling experimental measurements. Our results highlight the superior performance of the DP-SCAN and DP-LDA in accurately predicting high-temperature equations of states and elastic properties. This hybrid computational approach offers a solution to the accuracy-efficiency dilemma in obtaining precise elastic properties at high pressure and temperature conditions for minerals like MgPv, which opens a new way to study the Earth's interior state and related processes. 

\end{abstract}

\maketitle

\section{I. Introduction}
The Earth's lower mantle (LM), spanning from 670 to 2900 km in depth, constitutes approximately 60\% of the Earth's total volume. It is the largest continuous region of the planet, experiencing a wide range of thermodynamic conditions from around 23 GPa and ~1900 K to potentially 135 GPa and ~4000 K \cite{1}. The high-pressure Pbnm-perovskite polymorph of MgSiO$_3$ , i.e., bridgmanite (\textit{Bm})\cite{2}, is the most abundant\cite{3}. The physical properties of \textit{Bm}, including its structural and elastic properties, are of paramount geophysical interest; they underpin the overall properties of the lower mantle. High-temperature elasticity is a fundamental property of solids in geophysics, as it can be used to determine the speeds of seismic waves. 

To determine its pressure, volume, and temperature relationship, measurements of \textit{Bm}’s structural properties under extreme conditions have extensively utilized X-ray diffraction on samples compressed in laser-heated diamond anvil cells\cite{4,5,6}.  However, the full elastic tensor of \textit{Bm} at ambient conditions has only been determined experimentally on single-crystal samples\cite{4,5}, because direct measurement of the elastic moduli at lower mantle conditions poses significant challenges\cite{6}. Extensive extrapolations encompassing the entire P-T range relevant to the lower mantle (i.e., 23-135 GPa, \~1900-4000 K) can introduce additional uncertainty due to different physical models.

Given that the extreme pressure and temperature conditions within the Earth's deep interior are challenging to reach experimentally, insights from \textit{ab initio}-based methods have become essential. \textit{Ab initio} calculations have been extensively employed to verify and complement experimental studies of thermodynamic and elastic properties, phase stability, and effects of anharmonicity via the calculation of the phonon dispersion \cite{7,8,9,10,11,12,13,14,15}. Alternatively, molecular dynamics (MD) address full anharmonic effects and has proved accurate in obtaining elastic properties at sufficiently high temperatures. However, such simulations faced limitations of system sizes and simulation timescales due to the expensive computational costs and poor scaling of purely \textit{ab initio} calculations\cite{16,17}. Classical MD, while efficient, relies on the accuracy of conventional empirical potentials that sometimes fail to capture interatomic interactions over a wide range of pressures and temperatures encountered in the lower mantle.

Recently, machine learning methods have presented a solution to the accuracy-efficiency dilemma\cite{18,19,20,21} and have been successfully applied in MD simulations under extreme conditions. A few benchmarks have demonstrated that deep-learning potentials (DP)\cite{22} can achieve high accuracy in terms of force, energy, and the characterization of solid and liquid structures using hundreds to a few thousand reference configurations\cite{23,24,25}. The combination of deep-learning potentials with advanced density functionals, such as strongly constrained and appropriately normed (SCAN) meta-GGA functional\cite{26}, provides an opportunity to describe better the equation of states (EOS) than the traditional LDA or GGA functionals and overcome the multiplied computational costs over the conventional calculations.\cite{27}

In this study, we develop deep neural network potential models for \textit{Bm} using density functional theory (DFT) with various functionals. These potentials allow us to conduct extensive MD simulations covering a wide range of pressure-temperature (P-T) conditions, enabling a detailed investigation of the compressional behavior and elastic moduli of \textit{Bm} at high PTs. We compare the potential against pure DFT calculations and experimental measurements to evaluate its accuracy.

\section{II. Methods}
\subsection{2.1 DFT calculation}
To generate the \textit{ab initio} benchmark datasets for MgPv, we performed \textit{ab initio} MD simulations employing the Vienna \textit{ab initio} simulation package (VASP)\cite{28}. We employed multiple exchange-correlation functionals to model the system, including the local density approximation (LDA)\cite{29}, the Perdew-Burke-Ernzerhof parametrization (PBE) generalized gradient approximation\cite{30}, the revised PBE for solids (PBEsol)\cite{31}, and the SCAN functional\cite{26}. The projector augmented wave pseudopotential\cite{32,33} was employed, along with a plane-wave cutoff energy of up to 100 Ry. In AIMD simulations, we used 160-atom supercells of MgSiO$_3$ with $\Gamma$ \textbf{k}-point sampling. These DFT settings warranted good convergence of the calculated results\cite{34}.

\subsection{2.2 Development of Machine Learning Potentials}
Using the \textit{ab initio} datasets of forces and energies, we developed deep potential (DP) models for \textit{BM} employing the DeePMD-kit package\cite{35,36}. Two-body embedding with coordinates of the neighboring atoms (se\_e2\_a) was used for the descriptor. The embedding network was designed with a shape of (25, 50, 100), while the fitting network had a shape of (240, 240, 240). We used a cutoff radius of 6 Å and a smoothing parameter of 0.5 Å. The model was trained using the Adam optimizer\cite{37} for $1\times 10^6$ training steps, with the learning rate exponentially decaying from $1\times 10^{-3}$ to $3.51\times 10^{-8}$  during the training process. The loss function $\mathcal{L}(p_e,p_f)$ is given by\cite{35}: 
\begin{equation}
\mathcal{L}(p_e,p_f)=p_e\mid \Delta e \mid^2+\frac{p_f}{3N}\mid \Delta f_i \mid^2, \label{1}
\end{equation}
where $p_e$ linearly decays from 1.00 to 0.02, while $p_f$ linearly increases from $1\times 10^0$ to $1\times 10^{3}$ throughout the training process.

We employed the DP-GEN concurrent learning scheme to create the reference dataset and generate the potential\cite{38}. Initially, we randomly extracted 100 labeled configurations from 25 MD runs spanning various P-T ranges, covering 0-160 GPa and 300-4000 K to generate the initial potentials, shown in Fig.~\ref{fig:figS1}. We performed four DP-GEN iterations to explore the configuration space and ultimately achieve a potential that meets the desired accuracy for MD simulations. We trained four candidate DP potentials initialized with different random seeds in each iteration. The error estimator (model deviation) $\epsilon_t$ is determined based on the force disagreement between the candidate DPs\cite{38,39}. The expression for $\epsilon_t$ is: 
\begin{equation}
\epsilon_t\ =\ \mathop{max} \limits_{i} \sqrt{\left \langle \mid \mid F_{\omega,i}(\mathcal{R}_t) - \left \langle F_{\omega,i} (\mathcal{R}_t)\right \rangle \mid \mid^2 \right \rangle}
\end{equation}

where $F_{\omega,i}(\mathcal{R}_t)$ represents the force on the $i$-th atom predicted by the $\omega$-th potential for the configuration $\mathcal{R}_t$. After the DP-GEN iterations, the final training dataset comprises 3,000 configurations annotated with \textit{ab initio} force and energy information. We trained four potentials to reproduce the results of four exchange-correlation functionals: LDA, PBE, PBEsol, and SCAN. Consequently, we generated DP-LDA, DP-PBE, DP-PBEsol, and DP-SCAN potentials.

We used the LAMMPS package\cite{35} to perform MD simulations. We obtained phonon dispersions using the finite-displacement method implemented in Phonopy\cite{40} and PhonoLammps codes\cite{41} using a supercell of $2\times 2\times 2$ with 160 atoms. The elastic properties were determined with DPMD using stress-strain relation\cite{42,43,44} with 1,280 atoms and infinitesimal deformation.

\begin{figure}
\includegraphics[width=0.5\textwidth]{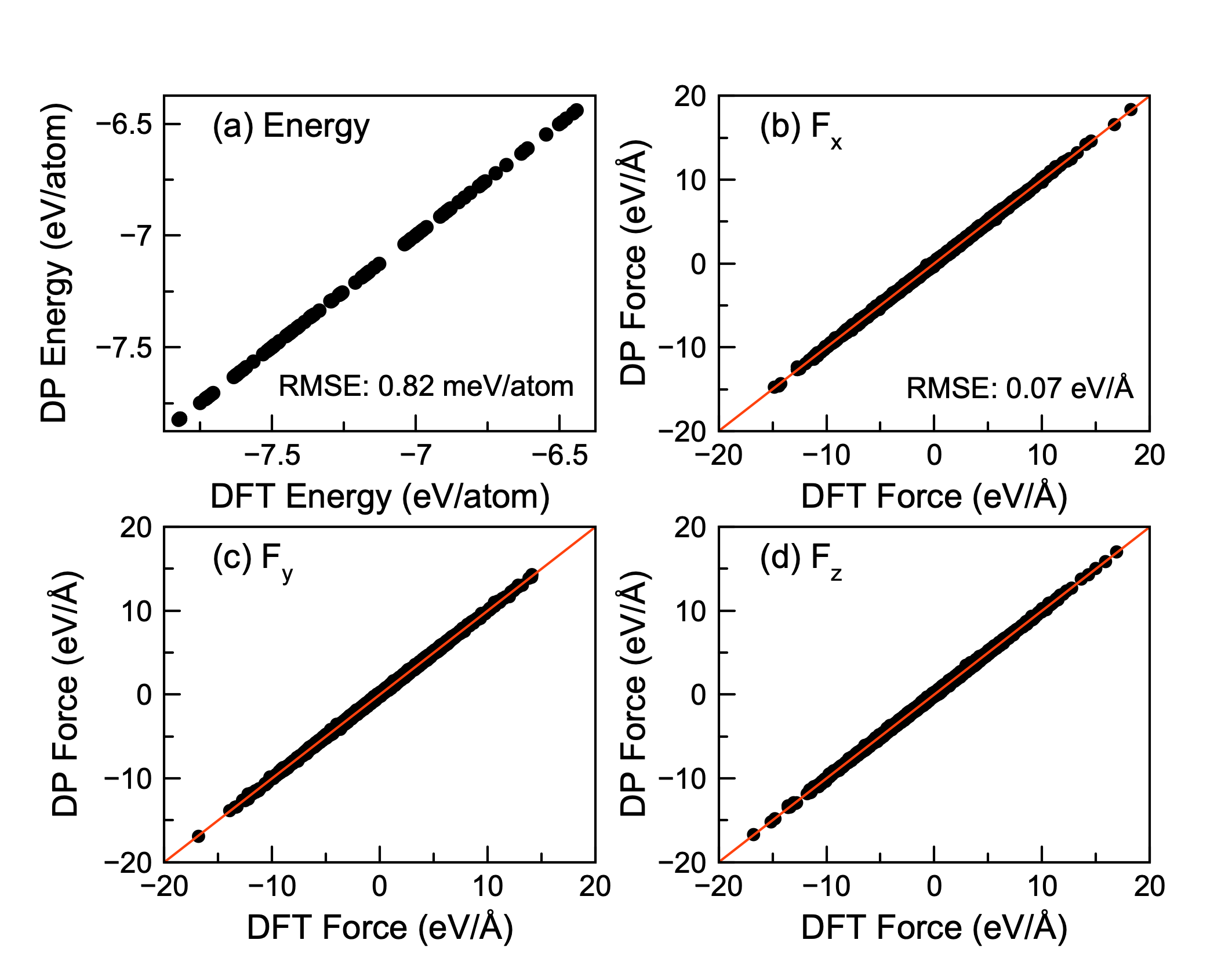}
\caption{\label{fig:fig1} Comparison of DP-LDA's and \textit{ab initio} calculations on the (a) potential energy and (b) (c) (d) atomic forces, which covers 0 - 160 GPa and 300 - 4000 K.}
\end{figure}

\section{III. Results}
\subsection{3.1 Validation of machine learning potential}
We first assess the accuracy of our machine learning potential. A comprehensive comparison between the DP and DFT calculations is undertaken for energies and atomic forces. We quantified the root-mean-square error (RMSE) for the DP's predictions on the validation sets. Fig.~\ref{fig:fig1} shows the outcomes of this analysis for the DP-LDA calculation. The RMSE for potential energies amounted to approximately 0.82 $meV/atom$, while the RMSE for atomic force prediction errors was around 0.07 $eV/ \AA$. Fig.~\ref{fig:figS2}-Fig.~\ref{fig:figS4} validate the results of force and energy for DP-PBE, DP-PBEsol, and DP-SCAN, respectively. All these potentials show small RMSE compared to the DFT data, indicating the present deep potentials are well trained.

\begin{figure}[t]
\includegraphics[width=0.5\textwidth]{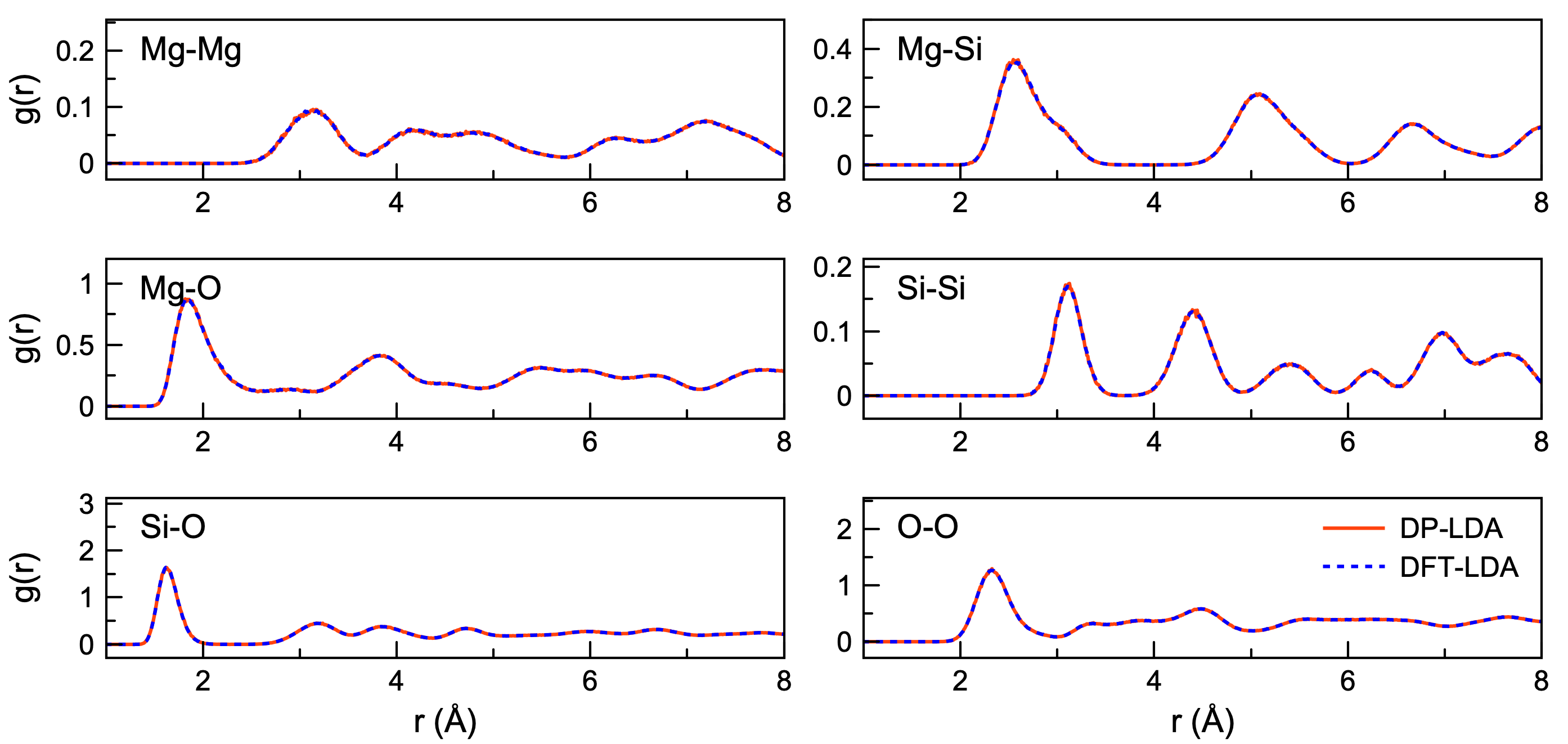}
\caption{\label{fig:fig2} Radial distribution functions of MgPv at 4000 K and 120 GPa from DFT-LDA calculations (dotted blue lines) and DP-LDA (solid red lines).}
\end{figure}

Fig.~\ref{fig:fig2} shows the radial distribution functions of \textit{Bm} at 4000 K and 120 GPa, obtained from both DFT-LDA and DP-LDA simulations. The DP potentials faithfully reproduce the $g(r)$ data derived from DFT calculations, indicating the DP's capability to capture \textit{Bm}’s structural and bonding properties accurately.

\begin{figure}
\includegraphics[width=0.48\textwidth]{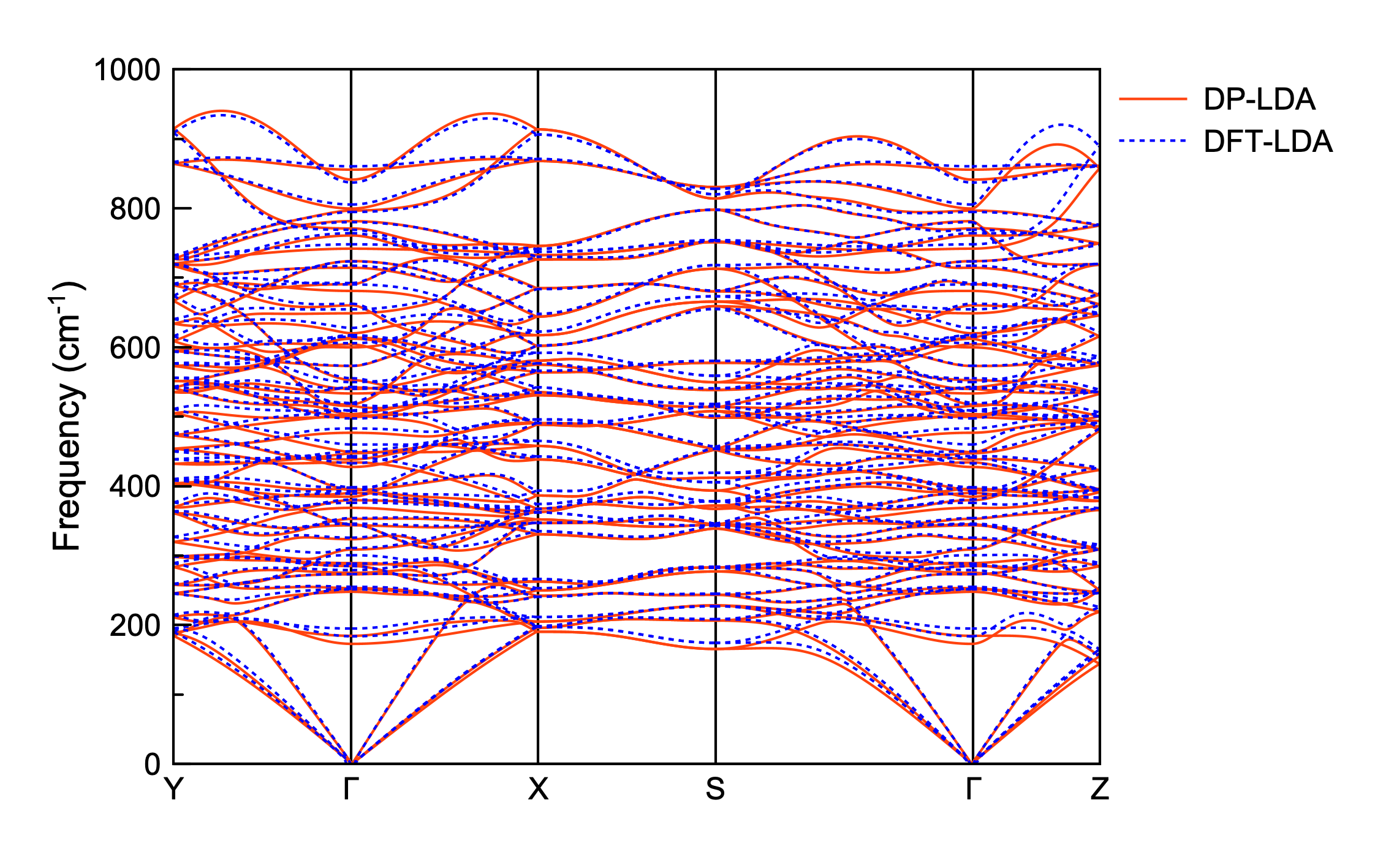}
\caption{\label{fig:fig3} Phonon dispersions of MgPv at 0 GPa calculated by DFT-LDA (dotted blue lines) and DP-LDA (solid red lines). }
\end{figure}

Since phonon dispersions are crucial in determining high-temperature properties, we also compare in Fig.~\ref{fig:fig3} these dispersions at 0 GPa obtained with DP-LDA and DFT-LDA calculations. We see excellent agreement across almost all phonon branches, with only minor deviations in a few high-frequency optical branches along $\Gamma-Z$ path. The agreement between DPMD and DFT predictions, encompassing force, potential energy, $g(r)$, and phonon dispersion, attest to the robustness of our DP potential. Therefore, we can confidently proceed and compute the EOS and elastic coefficients, $\textbf{c}_{ij}$, at high-pressure and high-temperature conditions.

\subsection{3.2 Equation of state}
\begin{figure}[t]
\includegraphics[width=0.45\textwidth]{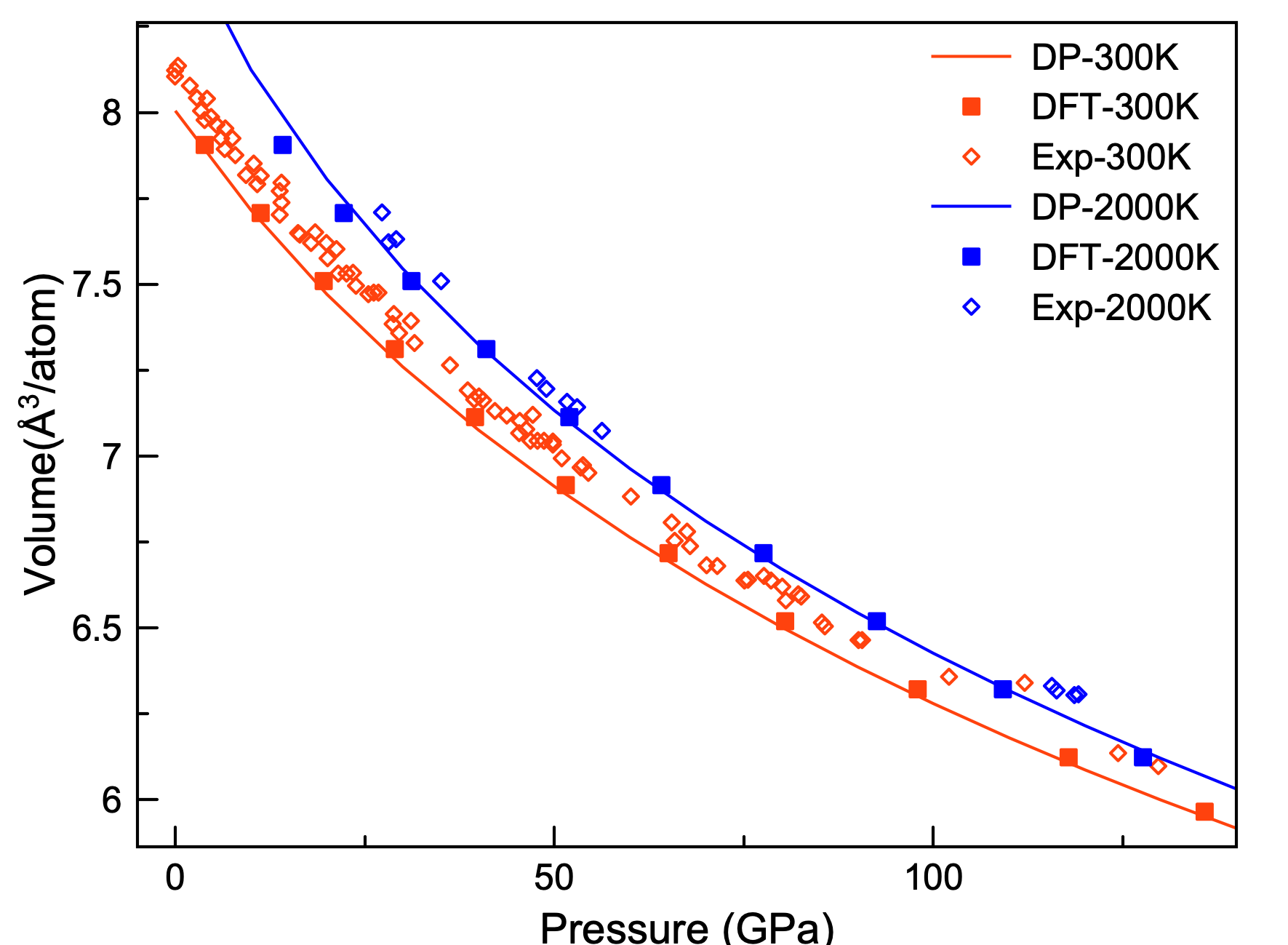}
\caption{\label{fig:fig4} Equation of state of MgPv at 300 K and 2000 K. Solid curves are obtained using DP-LDA, solid squares are the DFT-LDA results\cite{34}, and experimental data are taken from Refs.\cite{45,46,47,48,49,50,51,52,53,54}.}
\end{figure}

We compute the EOS across a broad range of P-T conditions. Fig.~\ref{fig:fig4} compares EOSs obtained from our DP and previous LDA calculations\cite{34}. Notably, the third-order Birch-Murnaghan EOS derived from DPMD results agrees exceedingly well with the DFT calculations. However, it is worth noting that both DPMD and DFT calculations tend to underestimate the pressure at a given volume when compared to experimental values\cite{45,46,47,48,49,50,51,52,53,54}. This discrepancy can be attributed mainly to the LDA functional employed in the present calculations. We also note that the DP calculations are based on classical MD simulations without taking into account the quantum effect of zero-point motion (ZPM). This ZPM effect was included in the previous LDA calculation\cite{34}, which manifests in the slight underestimation of DP-300 K compared with the DFT-300 K results. This ZPM effect is clearly reduced at 2000 K as the system approaches the classical limit. 

\begin{figure}
\includegraphics[width=0.48\textwidth]{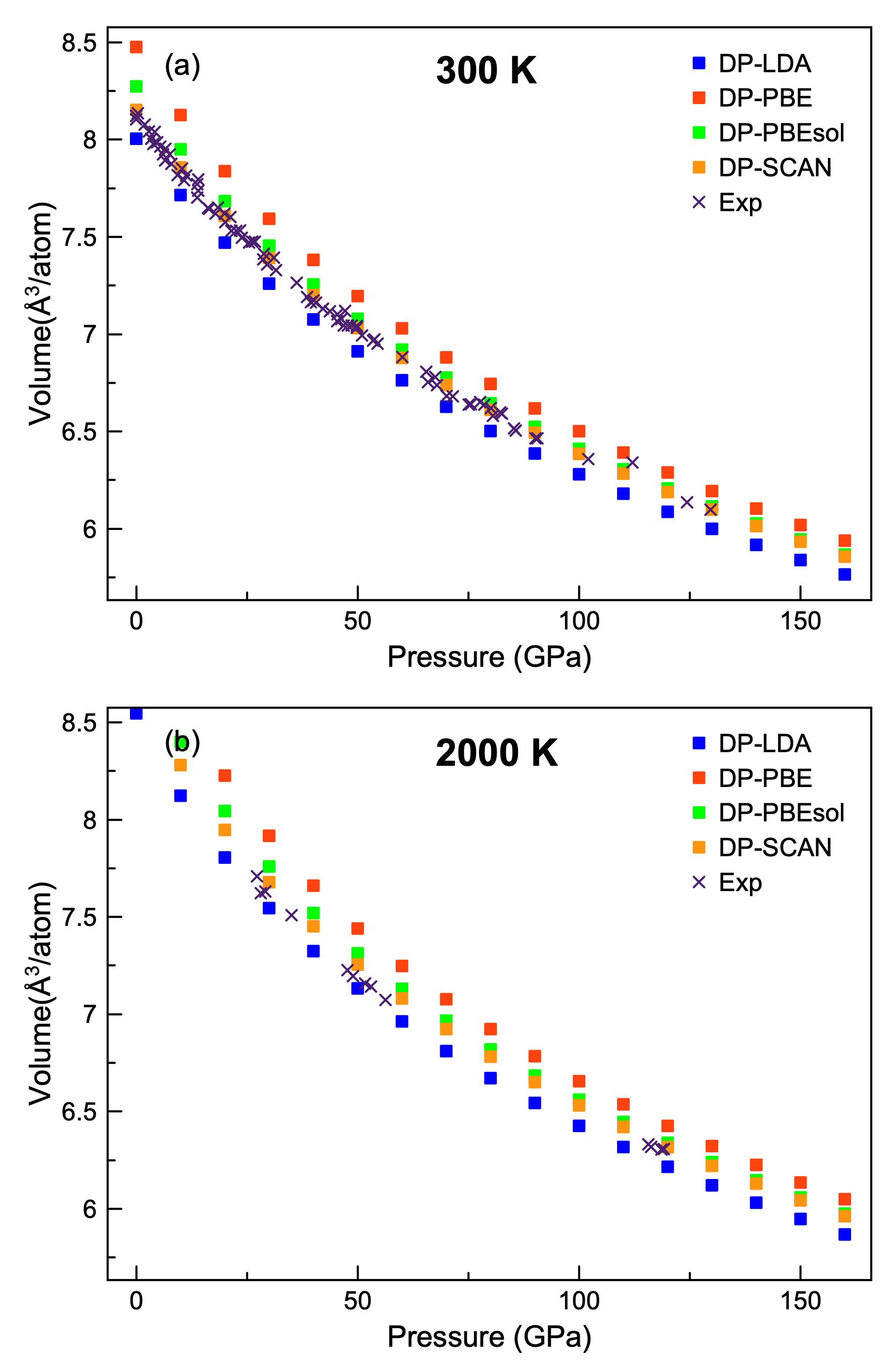}
\caption{\label{fig:fig5} Equation of state of MgPv at (a) 300 K and (b) 2000 K. Results from DP-LDA, DP-PBE, DP-PBEsol, and DP-SCAN correspond with blue, red, green, and orange squares, respectively. Experimental data are from Refs.\cite{45,46,47,48,49,50,51,52,53,54}.}
\end{figure}

Fig.~\ref{fig:fig5} compares the EOS obtained with various functionals and experimental data at 300 K and 2000 K. The commonly used DP-PBE and DP-PBEsol functionals tend to overestimate the volume; DP-LDA slightly underestimates it, while DP-SCAN performs significantly better. 

\subsection{3.3 Thermoelasticity}
\begin{figure}[t]
\includegraphics[width=0.48\textwidth]{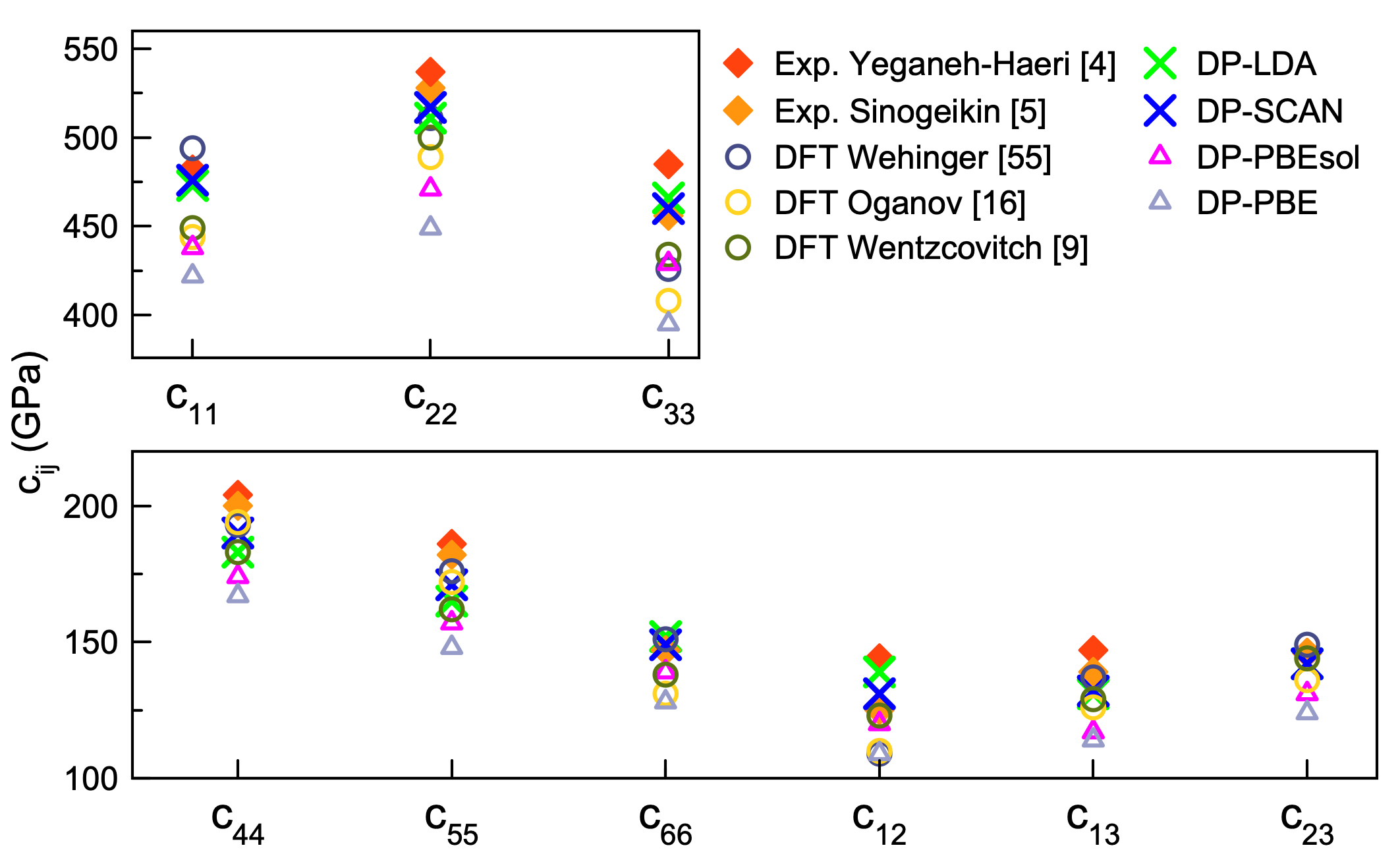}
\caption{\label{fig:fig6} Elastic moduli of Bm at 300 K and 0 GPa. Experimental values are from Ref.\cite{4,5}. Green and yellow circles are from previous GGA calculations\cite{16,55}, and brown circles are from previous LDA calculations\cite{9}. Crosses and up triangles are our DP results with different functionals.}
\end{figure}

We first compare in Fig.~\ref{fig:fig6} the elastic coefficients calculated using DPMD with those from previous PBE/GGA (DFT1, DFT2)\cite{16,55}, LDA (DFT3) calculations\cite{9}, as well as experimental measurement\cite{4,5} at 300 K and 0 GPa; a comprehensive list of these coefficients is provided in Table S1. One can see DP-SCAN yields results more consistent with experimental data, particularly $\textbf{c}_{11}$, $\textbf{c}_{22}$, and $\textbf{c}_{33}$. This underscores DP-SCAN's superior predictive capability for elastic properties. DP-LDA also agrees well with the experimental data, with minor discrepancies observed in $\textbf{c}_{44}$ and $\textbf{c}_{55}$. The DP-PBE and DP-PBEsol methods cannot accurately describe the elastic properties. Their deviation from the experimental data is approximately 12.5\%. It is noteworthy that previous DFT calculations yield different elastic properties at room temperature, particularly for $\textbf{c}_{11}$ by Wehinger et al.\cite{55}. Acoording to Wehinger et al.\cite{55}, this discrepancy was primarily attributed to previous calculations directly deriving results from acoustic phonon dispersions, bypassing numerical challenges arising from finite grid sampling in reciprocal space and a limited number of plane waves. In contrast, the present study with DP calculations enables large-scale MD simulations to directly investigate the stress-strain relationship and obtain accurate elastic behavior, which should ensure the reliability and robustness of our results.

Fig.~\ref{fig:fig7} depicts the dependence of the elastic coefficients on P and T, revealing an almost linear increase with P and a linear decrease with T. These results are consistent with previous measurements\cite{4} performed at ambient conditions. Furthermore, both the DP-SCAN and DP-LDA functionals agree well with experimental data at room temperature. Actually, for higher temperatures, our results closely match previous DFT calculations at 1500K\cite{16}. At 3500K, our results remain consistent with earlier ones for $\textbf{c}_{22}$\cite{16}, while showing minor discrepancies in $\textbf{c}_{11}$, $\textbf{c}_{44}$, and $\textbf{c}_{66}$. Since the present results fully address anharmonic effects and overcome the simulation size limitation, it is expected to provide more accurate results than previous ones. In principle, it requires high P-T experimental data to validate these predictions, but such measurements are unavailable. We note that the comparison with experimental data is based on the isothermal elastic coefficients, as the difference between isothermal and adiabatic elastic coefficients is relatively small. The inclusion of adiabatic correction may increse the non-shear elastic coefficients, particularly at high temperatures.\cite{13,56,57}

\textbf{
\begin{figure}[t]
\includegraphics[width=0.5\textwidth]{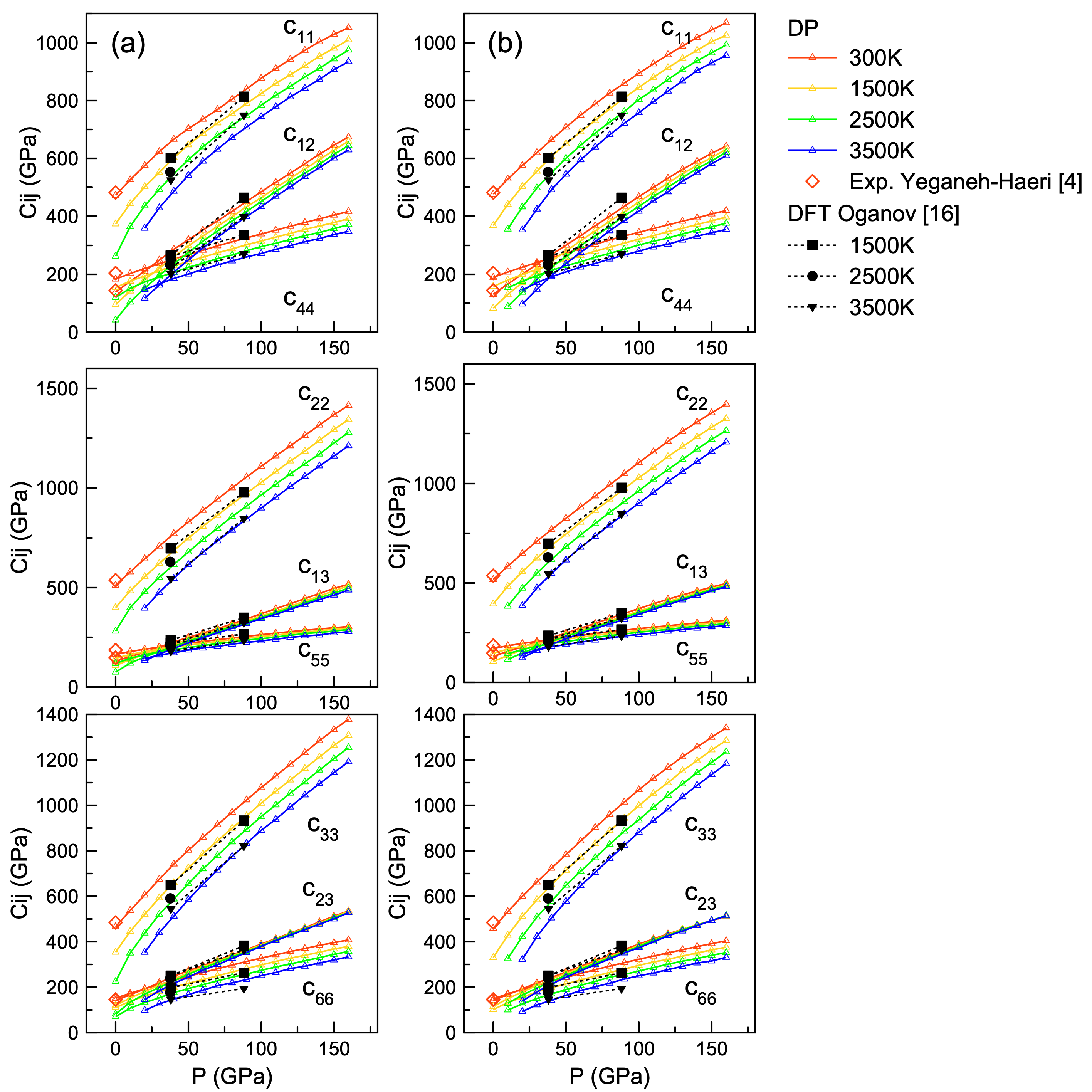}
\caption{\label{fig:fig7} Pressure dependence of the isothermal elastic constants of MgPv with (a) DP-LDA and (b) DP-SCAN. Solid lines correspond to results within our DP results. Open diamonds represent measurements at 0 Gpa\cite{4}. Square symbols are the results of Oganov et al.\cite{16} at 38 and 88 GPa. Symbol colors denote temperatures.}
\end{figure}
}

Fig.~\ref{fig:fig8} illustrates the pressure-temperature dependence of the Voigt-Reuss-Hill (VRH)-averaged\cite{58,59} bulk modulus ($K_S$), shear modulus ($G$), as well as the compressive ($V_p$) and shear ($V_S$) wave velocities from DP-LDA and DP-SCAN. These properties are derived from the elastic coefficients ($\textbf{c}_{ij}$) and exhibit a consistently positive pressure dependence and negative temperature dependence. In Fig.~\ref{fig:fig8} (a) and (b), we compare the predicted velocities and density of \textit{Bm} with seismic values from the Preliminary Reference Earth Model (PREM)\cite{60}. The density ($\rho$) is found to be relatively insensitive to temperatures, changing by only 1\% with a 1000 K variation at the conditions present at the bottom of the lower mantle. Both $V_{S}^{PREM}$ and $V_{p}^{PREM}$ closely match the 2500 K to 3500 K isothermal shear and compressive wave velocities, in the mantle. Moreover, $V_{p}^{PREM}$ intersects two $V_p$ isotherms, at 2500 K and 3500 K, suggesting a temperature difference of less than 1000 K from top to bottom in the lower mantle if it were entirely composed of \textit{BM}. Considering the shear and bulk moduli depicted in Fig.~\ref{fig:fig8} (c) and (d), our results agree well with experimental data at 300K. Both  $K_{S}^{PREM}$ and $G^{PREM}$  closely follow the corresponding 2500 K isotherms. Also, our shear modulus displays favorable agreement with previous DFT calculations\cite{16}, even at higher temperatures. Comparing (a) and (b), or (c) and (d), both DP-LDA and DP-SCAN can well describe these properties.

\textbf{
\begin{figure}[t]
\includegraphics[width=0.5\textwidth]{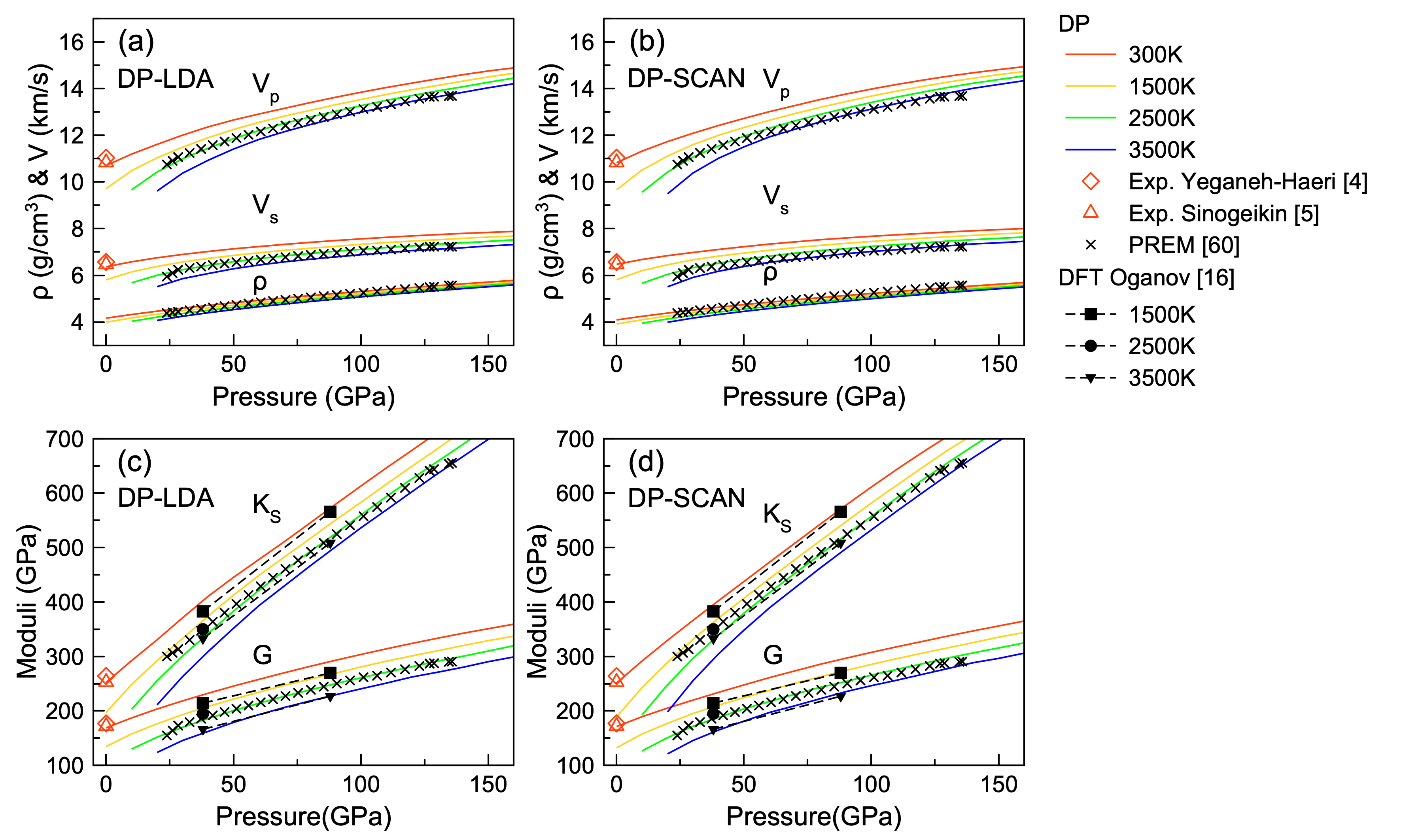}
\caption{\label{fig:fig8} Pressure dependence of density ($\rho$), isotropic longitudinal ($V_P=\sqrt{(K_S+4G/3)/\rho}$) and shear ($V_s=\sqrt{G/\rho}$) wave velocities, isothermal bulk moduli ($K_s$), and shear ($G$) moduli of MgPv. Solid lines in (a) and (c) are our DP-LDA results, while solid lines in (b) and (d) are our DP-SCAN results. Up and down triangles are the experimental results\cite{4,5}. Symbols connected by squares are the results of previous calculations\cite{16}. Colors denote temperatures. Black crosses are PREM data\cite{60}.}
\end{figure}
}

\section{IV. Discussion}
Our study highlights the accuracy of a hybrid approach that combines DFT with deep-learning (DP) potential to investigate the equation of state and elastic properties of \textit{Bm} at high PT conditions. The DP-GEN active learning scheme utilized to develop the potential demonstrates high efficiency, requiring only a few thousand reference DFT calculations completed within a few days. The computational cost of our DFT calculations is even lower than that of a typical AIMD run required for sampling a single (P, T) point. The resulting potential reproduces DFT results with excellent fidelity. This enhanced efficiency enables us to utilize more accurate functionals and conduct simulations on complex systems with improved accuracy in larger and more complex systems, longer timescales, and denser (P, T) sampling that were unattainable with pure DFT calculations. 
Our results indicate that the SCAN functional performs the best in calculations of the EOS and elastic properties of \textit{Bm}; the LDA functional performs nearly as well as the SCAN for elastic properties calculations, while PBE or PBEsol should be avoided. These results underscore the crucial importance of employing the most predictive functionals to accurately determine the thermodynamic properties of materials while emphasizing the potential for future high-PT research applications. 

The hybrid DFT-based machine learning approach transforms how we approach systems at extreme conditions with computational methods. It enables more accurate predictions for increasingly complex systems. Such an efficient and accurate approach for simulating the elastic properties of LM minerals at high PT conditions is poised to contribute to a deeper understanding of processes shaping the Earth's internal state.

\section{V. Conclusions}
Our study focuses on developing and testing deep learning potentials using different DFT functionals to investigate the EOS and thermoelasticity of \textit{Bm} under high-pressure and high-temperature conditions in the lower mantle. Our results demonstrate the accuracy of the DP approach by successfully reproducing radial distributions and phonon dispersions, with root mean square errors (RMSE) of 0.8 $meV/atom$ for energy and 0.07 $eV/\AA$ for atomic forces, further validating the robustness of our DP results. Regarding the EOS, we find that the choice of exchange-correlation functional in DFT calculations significantly impacts accuracy. DP-LDA slightly underestimates pressure at a given volume, while DP-PBE and DP-PBEsol overestimate it, even in the absence of ZPM effects. In contrast, the DP-SCAN method exhibits outstanding agreement with experimental data, demonstrating its superior capability for EOS prediction. Lastly, we establish the consistency of the elastic properties predicted by DP-SCAN and DP-LDA with previous quasiharmonic LDA calculations at high temperatures, confirming insignificant anharmonic effects on the high-temperature properties of \textit{Bm}. Additionally, mantle properties reported by the Preliminary Reference Earth Model (PREM) fall in the range of \textit{Bm} properties expected at lower mantle temperatures. We demonstrate that the synergistic application of DP and DFT provides a powerful means to predict EOS and elastic properties of mantle minerals more accurately at relevant temperatures, a task that has presented substantial experimental challenges.
\\\\[150pt]

\section{Acknowledgments}
Work at Columbia University was supported by the National Science Foundation awards EAR-1918126. T.W. and Y.S. acknowledge support from the Fundamental Research Funds for the Central Universities (20720230014). This work used Bridges-2 system at Pittsburgh Supercomputing Center, the Anvil system at Purdue University, the Expanse system at San Diego Supercomputing Center, and the Delta system at National Center for Supercomputing Applications from the Advanced Cyberinfrastructure Coordination Ecosystem: Services \& Support (ACCESS) program, which is supported by National Science Foundation grants \#2138259, \#2138286, \#2138307, \#2137603, and \#2138296. S. Fang and T. Wu from Information and Network Center of Xiamen University are acknowledged for the help with the GPU computing.

\twocolumngrid

\bibliographystyle{apsrev4-1}


\pagebreak
\widetext
\newpage

\setcounter{equation}{0}
\setcounter{figure}{0}
\setcounter{table}{0}
\renewcommand{\thefigure}{S\arabic{figure}}
\renewcommand{\thetable}{S\arabic{table}}

\begin{center}
\textbf{\Large{Supplementary Materials}}
\end{center}
\ \ \  \ \ \ \ \ \ \ \ \ \ \ \ \ \ \ \ \ \ \ \ \ \ \ \ \ \ \ \ \ \ \ \ \ \  This Supplementary Material includes: Figure S1-S4 and Table S1.

\begin{figure}[hp]
\includegraphics[width=0.8\textwidth]{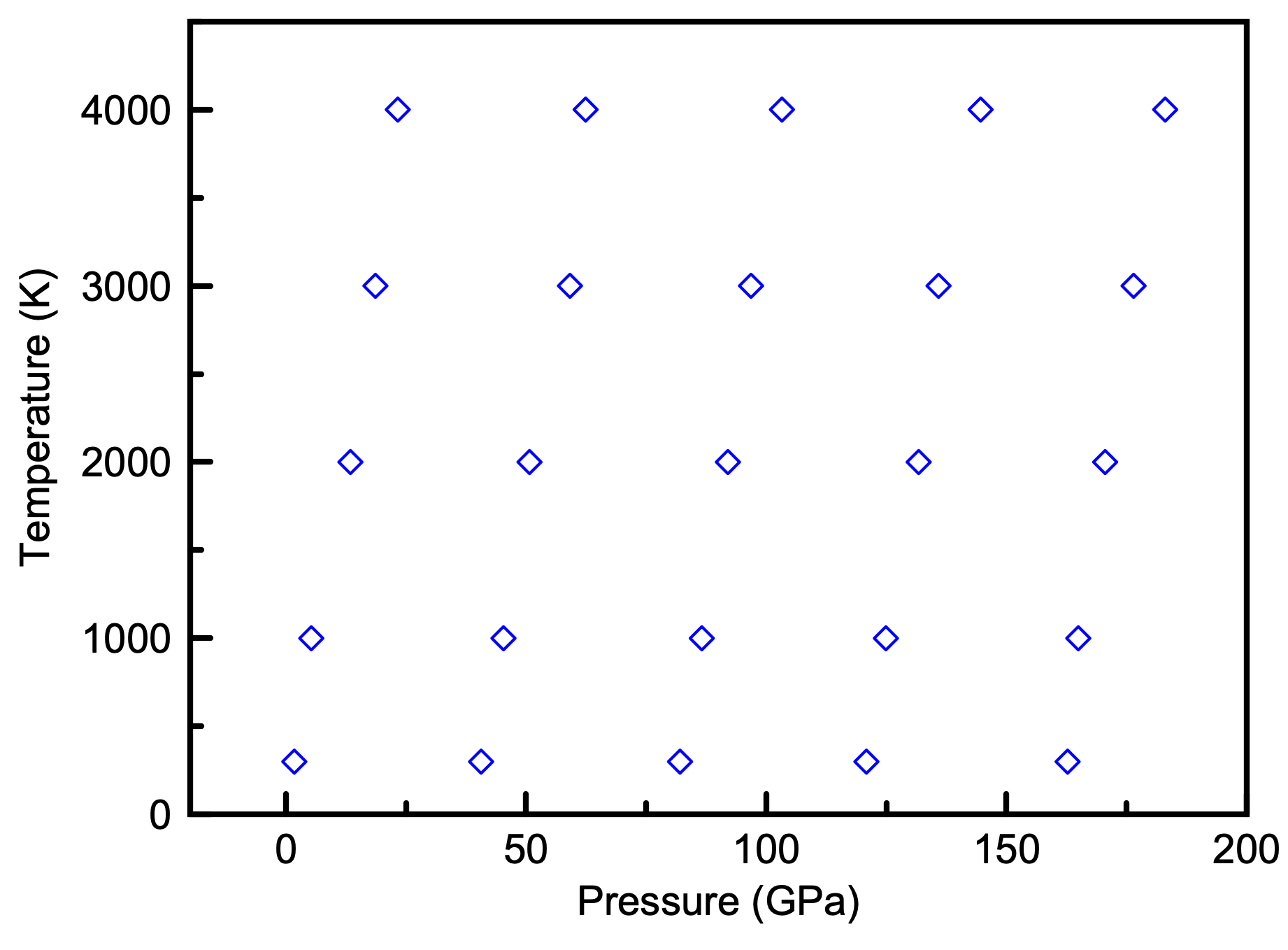}
\caption{\label{fig:figS1} P, T conditions covered by Born-Oppenheimer MD simulations.}
\end{figure}

\begin{figure}[hp]
\includegraphics[width=0.8\textwidth]{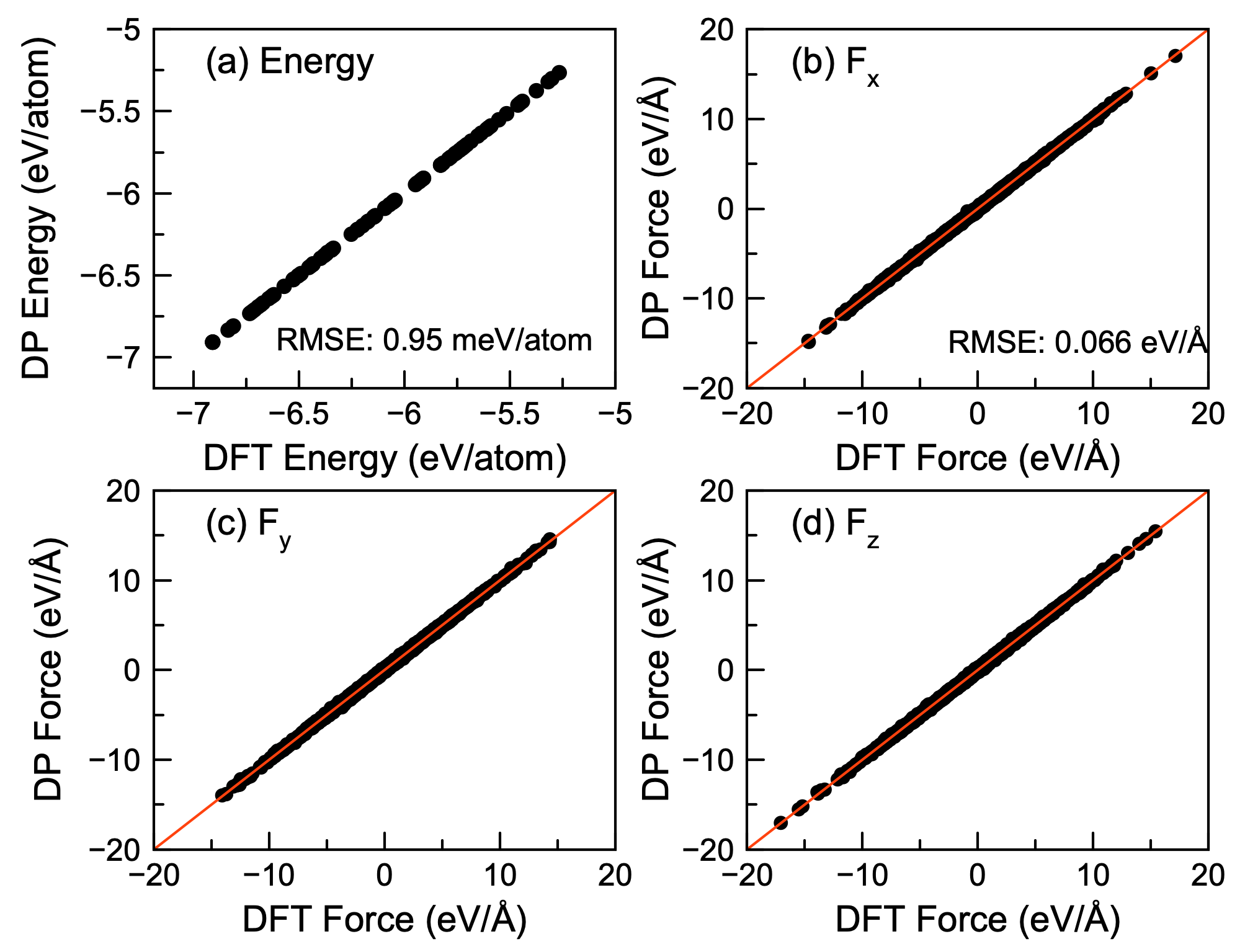}
\caption{\label{fig:figS2}Comparison of DP-PBE’s vs \textit{ab initio} prediction on the (a) potential energy and (b) (c) (d) atomic forces.}
\end{figure}

\begin{figure}[hp]
\includegraphics[width=0.8\textwidth]{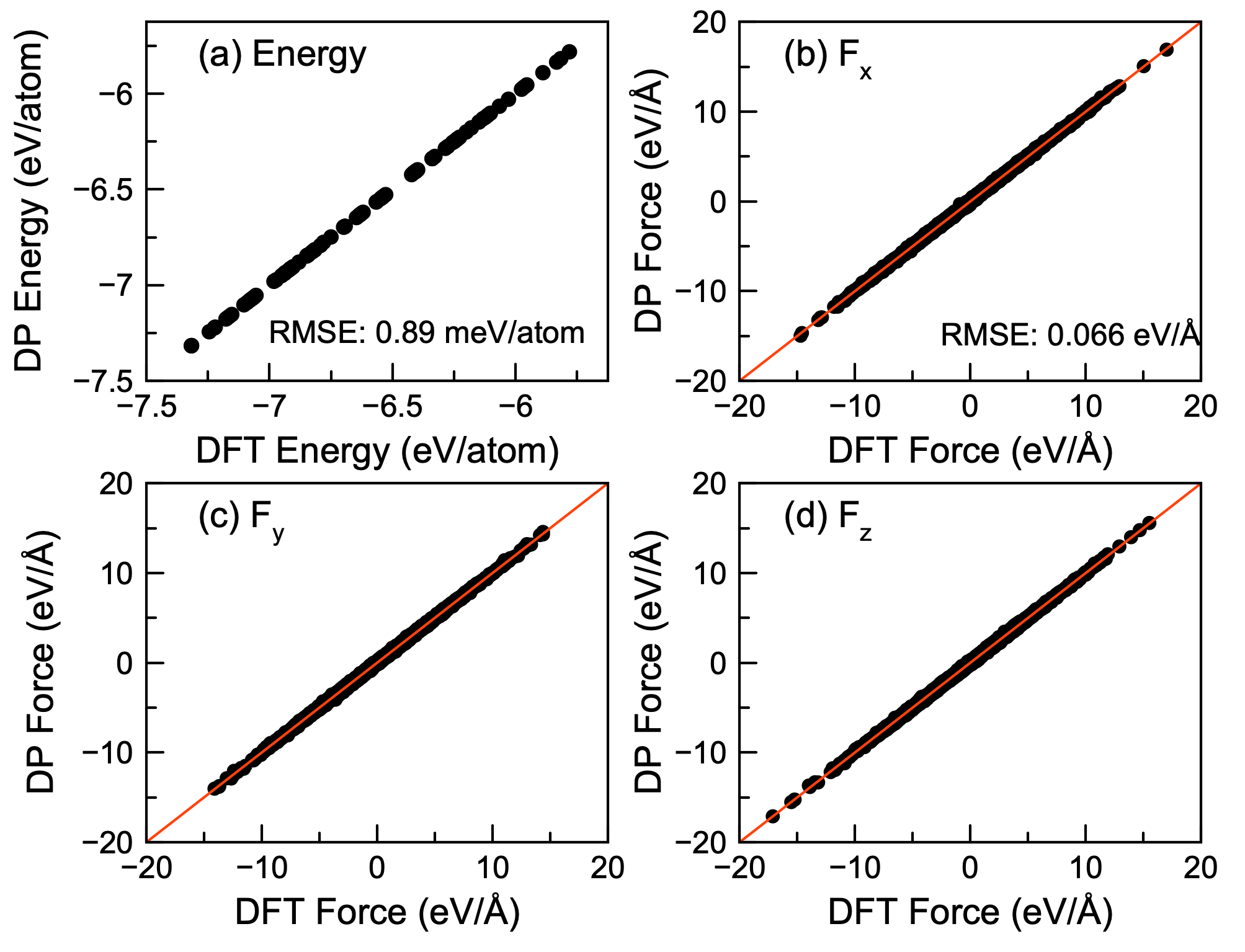}
\caption{\label{fig:figS3} Comparison of DP-PBEsol’s vs \textit{ab initio} prediction on the (a) potential energy and (b) (c) (d) atomic forces.}
\end{figure}

\begin{figure}[hp]
\includegraphics[width=0.8\textwidth]{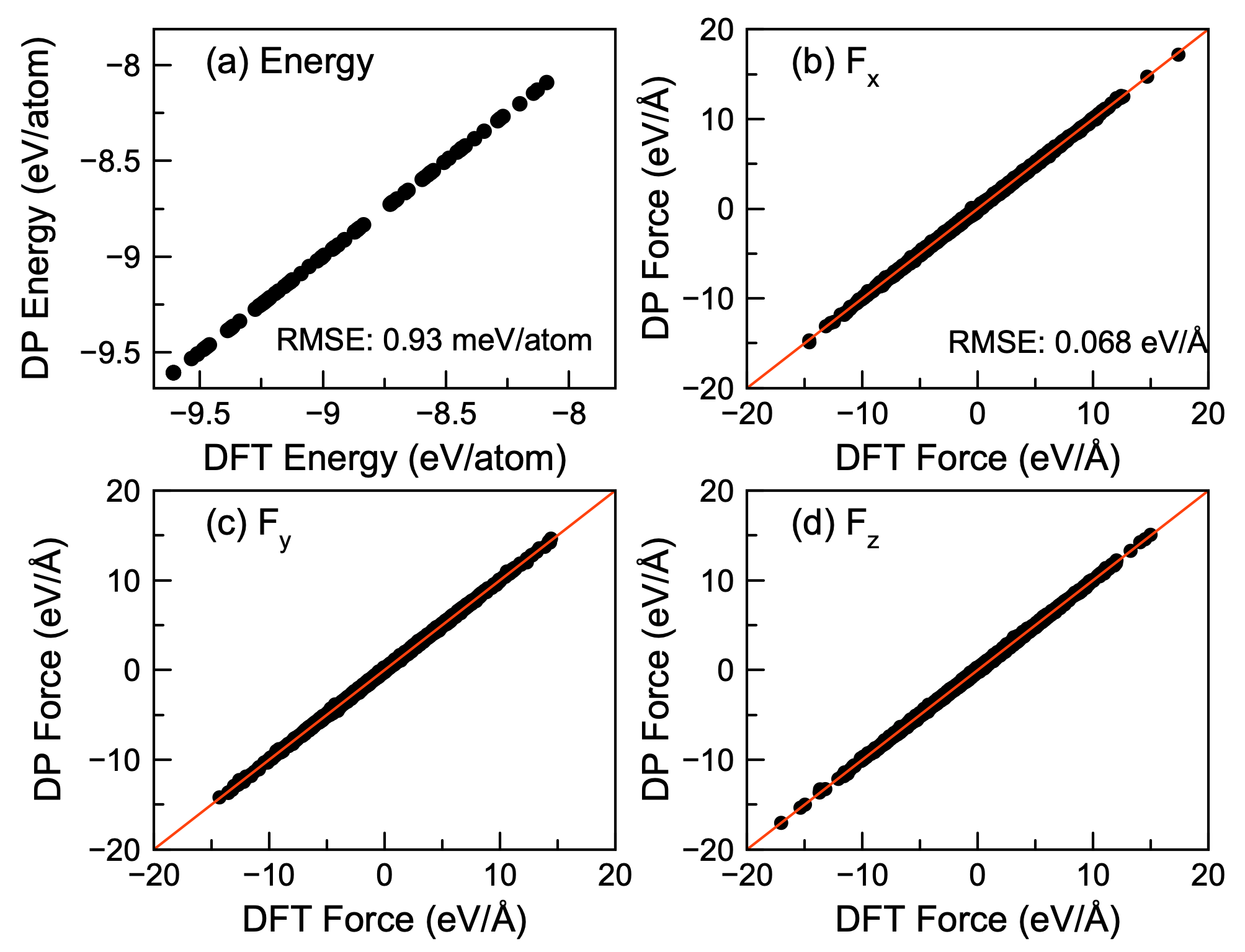}
\caption{\label{fig:figS4} Comparison of DP-SCAN’s vs \textit{ab initio} prediction on the (a) potential energy and (b) (c) (d) atomic forces.}
\end{figure}

\clearpage

\linespread{1.5}
\begin{table}[h]
    \centering
    \caption{Elastic moduli of MgPv at 300 K, 0 GPa. Experimental values are reported for comparison from Ref.\cite{1,2]}. Previous calculations are taken from Ref.\cite{3–5}.}

\begin{tabular}{p{3cm}<{\centering}p{1.5cm}<{\centering}p{1.5cm}<{\centering}p{1.5cm}<{\centering}p{1.5cm}<{\centering}p{1.5cm}<{\centering}p{1.5cm}<{\centering}p{1.5cm}<{\centering}p{1.5cm}<{\centering}p{1.5cm}<{\centering}}

\ &$\textbf{c}_{11}$&$\textbf{c}_{22}$&$\textbf{c}_{33}$&$\textbf{c}_{44}$&$\textbf{c}_{55}$&$\textbf{c}_{66}$&$\textbf{c}_{12}$&$\textbf{c}_{13}$&$\textbf{c}_{23}$\\
\hline EXP1\cite{1}&481&528&456&200&182&147&125&139&146\\
EXP2\cite{2}&482&537&485&204&186&147&144&147&146\\
GGA1\cite{3}&494&511&426&193&176&151&109&137&149\\
GGA2\cite{3}&440&488&412&183&158&133&114&123&135\\
GGA3\cite{4}&444&489&408&194&172&131&110&126&136\\
LDA1\cite{5}&449&500&434&183&162&138&123&129&144\\
DP-LDA&473&511&466&183&165&147&152&131&142\\
DP-SCAN&476&517&460&190&171&149&131&132&142\\
DP-PBEsol&438&471&429&174&157&139&120&117&131\\
DP-PBE&422&449&395&167&148&128&109&114&124\\

\end{tabular}

    \label{tab:my_label1}
\end{table}

\bibliographystyle{apsrev4-1}

\end{document}